\documentclass [submitting]{nst}

\usepackage{subfigure,dcolumn}
\usepackage[T2A,T1]{fontenc}
\usepackage[russian,english]{babel}
\usepackage{mathrsfs}
\usepackage{color,xcolor}

\usepackage{listings}
\begin {document}

\title{  Compact CubeSat Gamma-Ray Detector for GRID Mission}
\thanks{M. Zeng acknowledges funding support from the Tsinghua University Initiative Scientific Research Program. H. Feng acknowledges funding support from the National Natural Science Foundation of China (Grant Nos. 11633003, 12025301, and 11821303) and the National Key R\&D Program of China (Grant Nos. 2018YFA0404502 and 2016YFA040080X).}

\author{Jia-Xing Wen}
\affiliation{Key Laboratory of Particle and Radiation Imaging (Tsinghua University), Ministry of Education, Beijing 100084, China}
\affiliation{Department of Engineering Physics, Tsinghua University, Beijing 100084, China}
\affiliation{Science and Technology on Plasma Physics Laboratory, Laser Fusion Research Center, CAEP, Mianyang 621900, Sichuan, China}
\author{Xu-Tao Zheng}
\affiliation{Key Laboratory of Particle and Radiation Imaging (Tsinghua University), Ministry of Education, Beijing 100084, China}
\affiliation{Department of Engineering Physics, Tsinghua University, Beijing 100084, China}
\author{Jian-Dong Yu}
\affiliation{School of Electronic and Information Engineering, Ningbo University of Technology, Ningbo 315201, China}
\author{Yue-Peng Che}
\affiliation{School of Electronic and Information Engineering, Ningbo University of Technology, Ningbo 315201, China}
\author{Dong-Xin Yang}
\affiliation{Department of Astronomy, Tsinghua University, Beijing 100084, China}
\author{Huai-Zhong Gao}
\affiliation{Key Laboratory of Particle and Radiation Imaging (Tsinghua University), Ministry of Education, Beijing 100084, China}
\affiliation{Department of Engineering Physics, Tsinghua University, Beijing 100084, China}
\author{Yi-Fei Jin}
\affiliation{Key Laboratory of Particle and Radiation Imaging (Tsinghua University), Ministry of Education, Beijing 100084, China}
\affiliation{Department of Engineering Physics, Tsinghua University, Beijing 100084, China}
\author{Xiang-Yun Long}
\affiliation{Department of Astronomy, Tsinghua University, Beijing 100084, China}
\author{Yi-Hui Liu}
\affiliation{Department of Engineering Physics, Tsinghua University, Beijing 100084, China}
\author{Da-Cheng Xu}
\affiliation{Department of Engineering Physics, Tsinghua University, Beijing 100084, China}
\author{Yu-Chong Zhang}
\affiliation{Department of Physics, Tsinghua University, Beijing 100084, China}
\author{Ming Zeng}
\email[Corresponding author, ]{Ming Zeng, zengming@mail.tsinghua.edu.cn}
\affiliation{Key Laboratory of Particle and Radiation Imaging (Tsinghua University), Ministry of Education, Beijing 100084, China}
\affiliation{Department of Engineering Physics, Tsinghua University, Beijing 100084, China}
\author{Yang Tian}
\affiliation{Key Laboratory of Particle and Radiation Imaging (Tsinghua University), Ministry of Education, Beijing 100084, China}
\affiliation{Department of Engineering Physics, Tsinghua University, Beijing 100084, China}
\author{Hua Feng}
\affiliation{Key Laboratory of Particle and Radiation Imaging (Tsinghua University), Ministry of Education, Beijing 100084, China}
\affiliation{Department of Astronomy, Tsinghua University, Beijing 100084, China}
\author{Zhi Zeng}
\affiliation{Key Laboratory of Particle and Radiation Imaging (Tsinghua University), Ministry of Education, Beijing 100084, China}
\affiliation{Department of Engineering Physics, Tsinghua University, Beijing 100084, China}
\author{Ji-Rong Cang}
\email[Corresponding author, ]{Ji-Rong Cang, cangjr14@tsinghua.org.cn}
\affiliation{Key Laboratory of Particle and Radiation Imaging (Tsinghua University), Ministry of Education, Beijing 100084, China}
\affiliation{Department of Astronomy, Tsinghua University, Beijing 100084, China}
\affiliation{Department of Engineering Physics, Tsinghua University, Beijing 100084, China}
\author{Qiong Wu}
\affiliation{Key Laboratory of Particle and Radiation Imaging (Tsinghua University), Ministry of Education, Beijing 100084, China}
\affiliation{Department of Astronomy, Tsinghua University, Beijing 100084, China}
\author{Zong-Qing Zhao}
\affiliation{Science and Technology on Plasma Physics Laboratory, Laser Fusion Research Center, CAEP, Mianyang 621900, Sichuan, China}
\author{Bin-Bin Zhang}
\affiliation{Key Laboratory of Modern Astronomy and Astrophysics (Nanjing University), Ministry of Education, Nanjing 210093, Jiangsu, China}
\affiliation{Nanjing University, School of Astronomy and Space Sciences, Nanjing 210093, Jiangsu, China}
\author{Peng An}
\affiliation{School of Electronic and Information Engineering, Ningbo University of Technology, Ningbo 315201, China}
\author{GRID collaboration}

\begin{abstract}
Gamma-Ray Integrated Detectors (GRID) mission is a student project    designed to use multiple gamma-ray detectors carried by nanosatellites (CubeSats), forming a full-time all-sky gamma-ray detection network that monitors the transient gamma-ray sky in the multi-messenger astronomy era. A compact CubeSat gamma-ray detector, including its hardware and firmware, was designed and implemented for the mission. The detector employs four $\rm Gd_2Al_2Ga_3O_{12}:Ce$ (GAGG:Ce) scintillators coupled with four silicon photomultiplier (SiPM) arrays to achieve a high gamma-ray detection efficiency between $\rm 10\ keV$ and 2\ MeV with low power and small dimensions. The first detector designed by the undergraduate student team onboard a commercial CubeSat was launched into a Sun-synchronous orbit on October 29, 2018. The detector was in a normal observation state and accumulated data for approximately one month after on-orbit functional and performance tests, which were conducted in 2019.
\end{abstract}


\maketitle

\section{Introduction}
\label{sec:intro}

Gamma-Ray Integrated Detectors (GRID) mission is a student project designed for the scientific purpose of monitoring the transient gamma-ray sky in the local universe, particularly, to accumulate a sample of gamma-ray bursts (GRBs) associated with gravitational waves (GWs). According to the estimation of the GW-GRB joint detection event rate, the maximum number of events that can be detected is more than a dozen per year. Therefore, GRID is designed to serve as full-time all-sky gamma-ray detection networks, without Earth occultation or interruptions due to the South Atlantic Anomaly (SAA), with many compact and modularized gamma-ray detectors on a fleet of CubeSats in low Earth orbit. As a distributed system, GRID can localize detected GRBs via triangulation or flux modulation with simple scintillation detectors\cite{bib:1}.

The scientific payloads of GRID comprise several modularized compact gamma-ray detectors. Many key technologies have been utilized to optimize the gamma-ray detection performance in the limited space of a CubeSat. In the first gamma-ray detector prototype of GRID, GAGG:Ce was used, and the package was optimized for high transmittance of low-energy X-rays down to 10 keV. To respect the power and space limitations  of CubeSats, silicon photomultipliers (SiPMs) were utilized instead of traditional photomultiplier tubes (PMTs) owing to their attractive capabilities, such as their super miniature size, low weight, low power consumption, and insensitivity to magnetic fields\cite{bib:8}. The strong dark noise of the SiPM array restricts the signal-to-noise ratio (SNR) at room temperature; hence, the current sensitive pre-amplifier was designed, modeled, and optimized to improve the SNR. Data acquisition DAQ) electronics were designed based on an off-the-shelf ARM core microcontroller unit (MCU), which is sufficiently simple  for undergraduate students. We note that numerous CubeSat-based missions similar to GRID mission have been proposed and are under development\cite{bib:2}\cite{bib:3}\cite{bib:4}\cite{bib:5}\cite{bib:6}\cite{bib:7}. The scientific payloads of BurstCube\cite{bib:3}, CAMELOT\cite{bib:4}, HERMES\cite{bib:6}, and GRBAlpha (in-orbit demonstration for CAMELOT)\cite{bib:5} are scintillator-based detectors similar to those of GRIDs. BlackCAT\cite{bib:2} adopts silicon detectors and is sensitive to soft X-rays. LECX\cite{bib:7} is designed to employ four CdZnTe (CZT) detectors with high energy resolution.

The first detector prototype onboard a 6 U ($\rm 30\ cm \times 20\ cm \times 10\ cm$) CubeSat developed by Spacety Co.\ Ltd, a commercial satellite company in China, was launched into a Sun-synchronous orbit\cite{bib:1} with an altitude of $500$ km and inclination of $97.5^\circ$. It was in a normal observation state and accumulated data for approximately one month after its on-orbit functional and performance tests, which were conducted in 2019. Compared with the first detector, the second detector has fewer hardware modifications and improvements. An aluminized polyimide film was used to better insulate the sunlight, and the leakage current monitoring circuits of the SiPM arrays were modified to provide a wider measurement range in the second detector. The second detector was launched into a Sun-synchronous orbit with an altitude of $500$ km and inclination of $97.3^\circ$ by the Long-March 6 rocket on November 6, 2020. The second detector has accumulated data for more than 300 h of on-orbit observation. Multiple GRBs were observed. All the corresponding scientific data will be collected and published by the National Space Science Data Center (NSSDC) in the future. In this paper, we present the detailed design of the detector, electronics, and firmware. In addition, we discuss the energy resolution, low-energy X-ray detection performance, and high-rate performance of the detector.

\section{Detector structure}
\begin{figure}[!htb]
	\includegraphics[width=0.95\hsize]{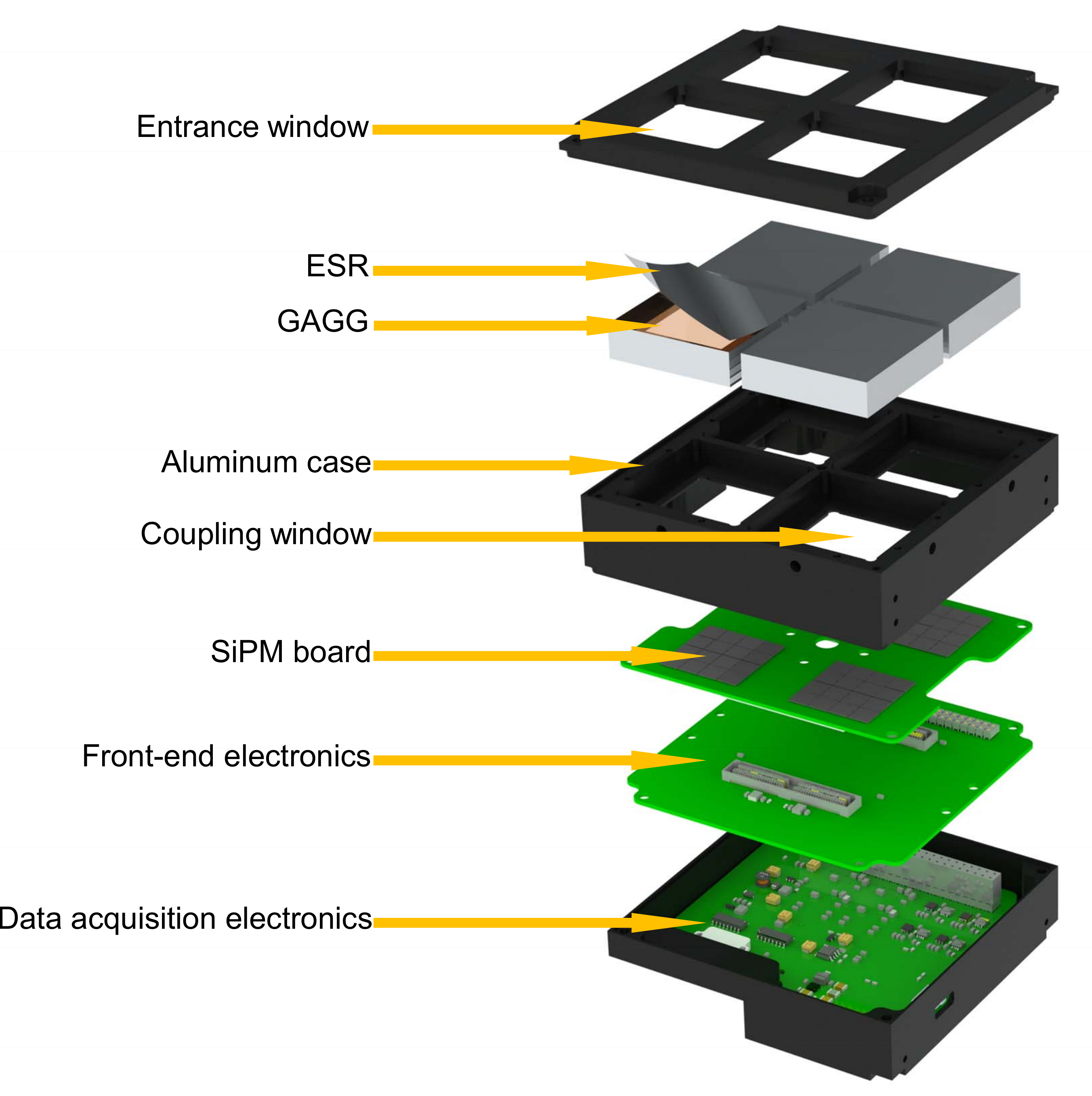}
	\includegraphics[width=0.7\hsize]{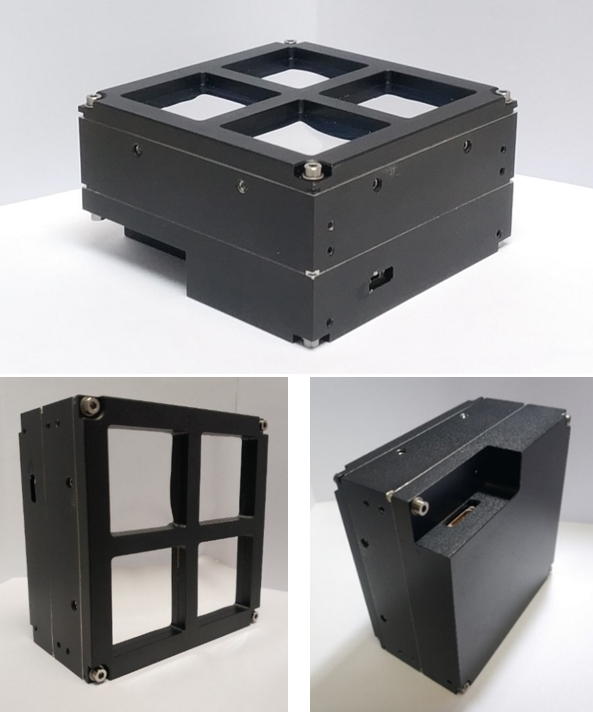}
	\caption{3D model of a GRID (top) and photograph of a GRID (bottom)}
\label{fig:detector_3D}
\end{figure}

A schematic and photograph of the first detector fabricated for GRID mission are shown in Fig.\ref{fig:detector_3D}. The detector comprises four GAGG:Ce scintillators coupled with four SiPM arrays on one SiPM board. Each SiPM array comprises $\rm 4\times4$ J-60035-type SiPMs from SensL. The standard output pins of the SiPMs are connected in parallel on a front-end electronics (FEE) board. The following DAQ board provides the capabilities of four-channel signal digitization, power distribution, communication, and detector control. The detector has dimensions of $\rm 5\ cm \times 9.4\ cm \times 9.4\ cm$, which occupies half of a CubeSat with standard dimensions (units or ‘U’), and a power consumption of 5 V$\times$0.6 A. The detector provides a novel electronic interface, supporting the serial peripheral interface (SPI), universal asynchronous receiver/transmitter (UART) protocols, and pulses per second (PPS) interface. The features of the first detector, which is a modularized CubeSat payload suitable for different nanosatellite platforms, are summarized in Table~\ref{table:features}.  
\begin{table}[!htb]
\caption{Main features of the first compact gamma-ray detector of GRID.}
\label{table:features}
\begin{tabular*}{8cm} {@{\extracolsep{\fill} } ll}
\toprule
Detector size & $\rm 0.5\ U\ (9.4\times9.4\times5\ cm^3$)\\ 
\hline
Weight & $\rm 780\ g$\\ 
\hline
Power & $\rm \leq3\ W $\\
\hline
Detection area & $\rm 58\ cm^2$\\
\hline
FOV & $\rm 1 \pi$\\
\hline
Energy range & $\rm 10\ keV \sim 2\ MeV$\\
\hline
Dead time & $\rm 50\ \mu s$ \\
\bottomrule
\end{tabular*}
\end{table}

\section{Scintillator and FEE}
\label{sec:det}
\subsection{Scintillator and package}
We used GAGG:Ce scintillators manufactured  by EPIC CRYSTAL Co.\ Ltd. Owing to the limited available size of GAGG:Ce scintillator, the detector comprises four GAGG:Ce scintillators to make full use of the area of a standard nanosatellite unit. A single GAGG:Ce scintillator has a surface area of $\rm 3.8\ cm\times3.8\ cm$ and a thickness of $\rm 1\ cm$. GAGG:Ce has the advantages of a high density ($\rm 6.5\ g/cm^3$) and a high detection efficiency for gamma-rays of up to the MeV magnitude. The maximum emission spectrum for GAGG:Ce is approximately 530 nm, which is suitable for silicon-based photodetectors. Its high light yield ($\rm \sim$30--70 ph/keV) and $\rm \sim 100\ ns$ decay time provide a reasonable SNR when the SiPM arrays are used. The reflection layer is an enhanced specular reflector (ESR), a 65 $\mu$m polymer with high reflectance ($\rm > 98\%$) manufactured by 3M. A series of aluminized polyester films, which are generally used as the outermost cover of satellites, is used as the light shield layer. Owing to the good mechanical characteristics and non-hygroscopic properties of GAGG:Ce\cite{bib:9}, no further layers are required to ensure sufficient moisture proofing and mechanical reinforcement. The simplified scintillator package provides a high detection efficiency of low-energy X-rays at a low cost.

\begin{figure}[!htb]
\includegraphics[width=0.35\hsize]{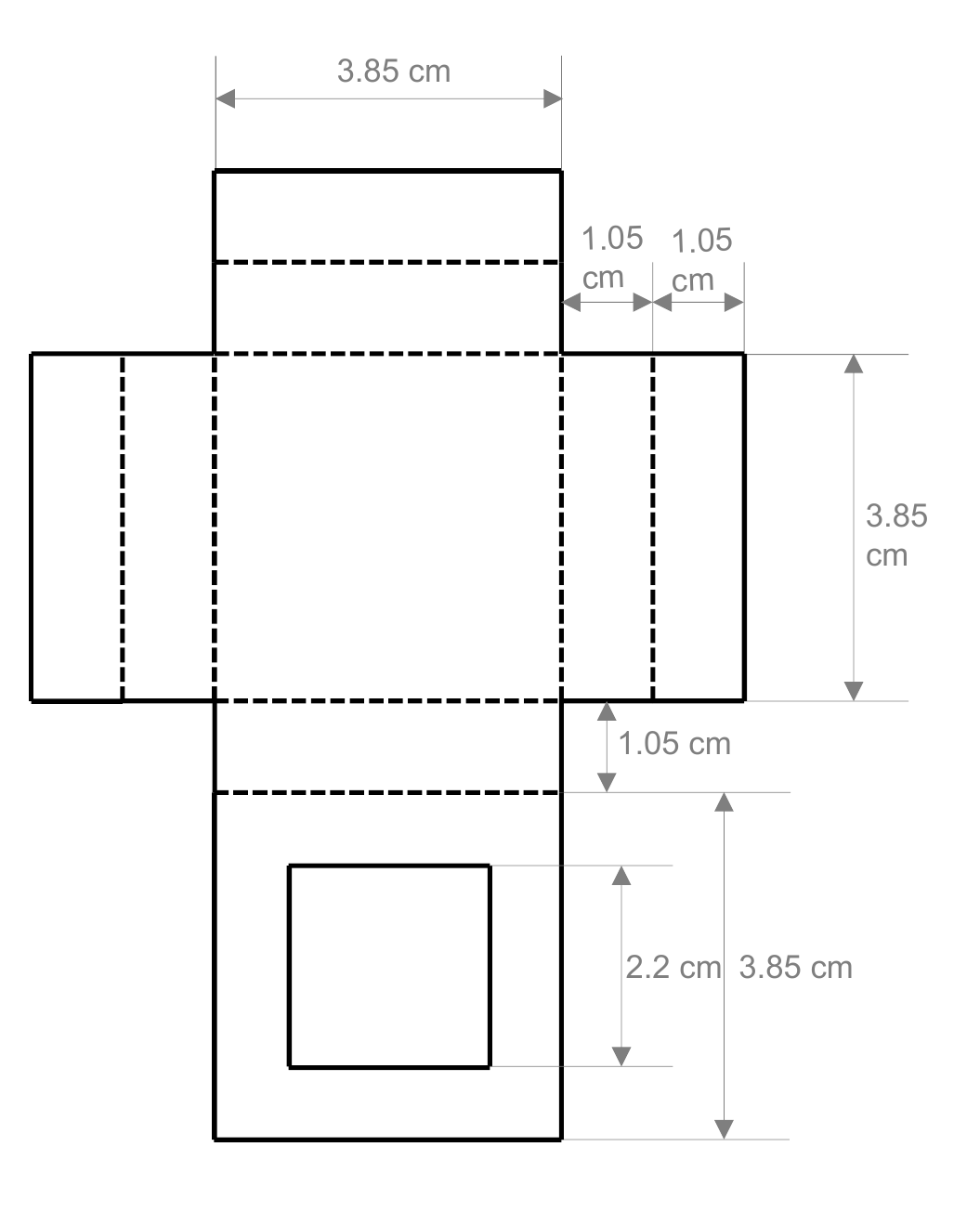}
\includegraphics[width=0.55\hsize]{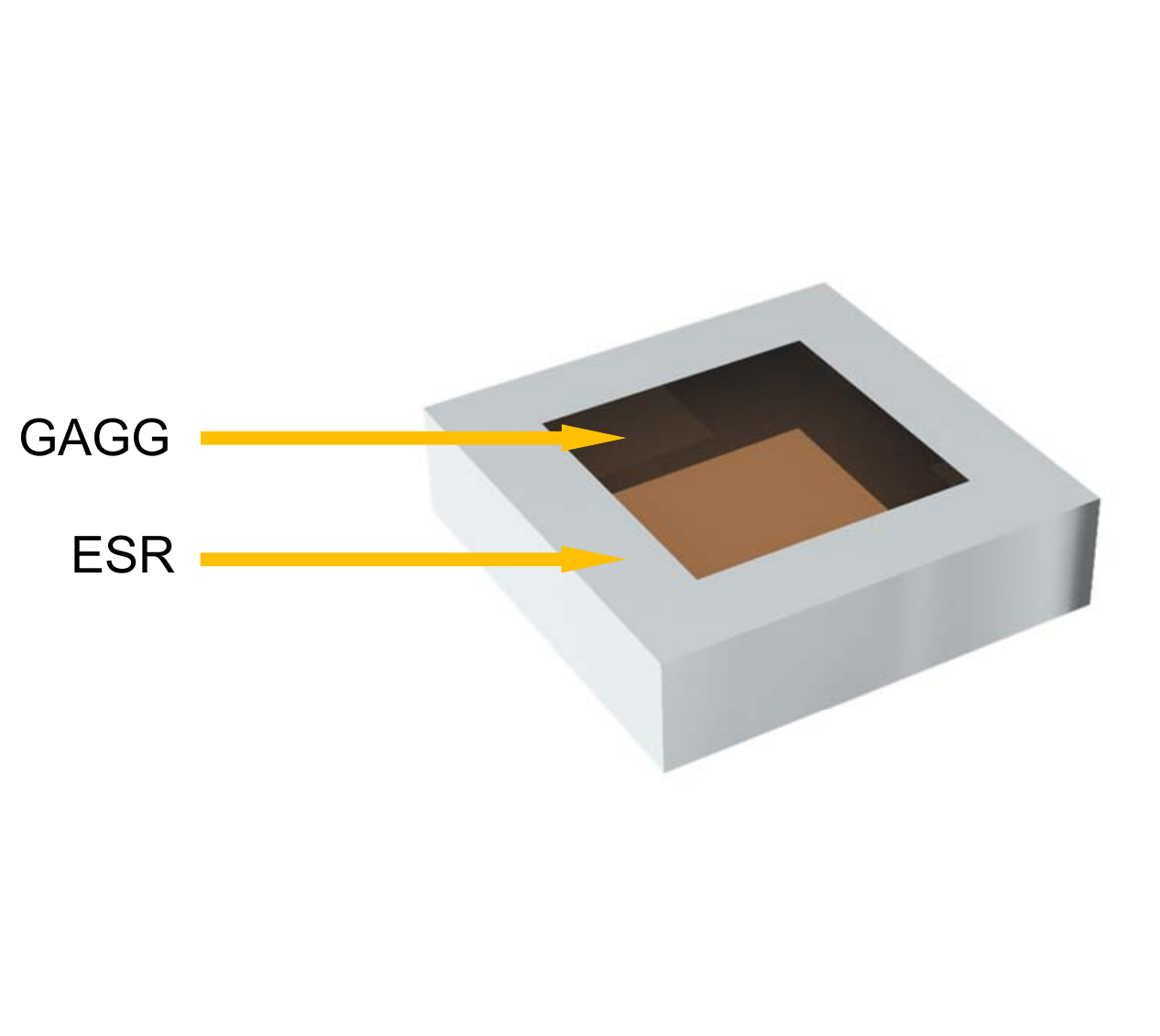}
\caption{ESR package (left), where the solid line is the outline, and the dashed line is the crease. Bottom view of a GAGG:Ce scintillator with the ESR package (right). The ESR is wrapped directly on the scintillator and taped on the side and bottom.}
\label{fig:ESRGAGG}
\end{figure}

With an increase in the number of parallel connected SiPMs on the same channel, the output capacitance and dark count rate of the SiPM array increase, which reduce the SNR. Therefore, an SiPM array smaller than the bottom area of the scintillator is utilized, similar to many detectors where the scintillator is coupled with SiPMs. The ESR is cut into a particular shape and fully covers  the top and side of the scintillator. Moreover, there is a $\rm 2.2\ cm\times2.2\ cm$ square window at the bottom-center for the scintillation light collection, as shown in Fig.\ref{fig:ESRGAGG}. By measuring the 661.7 keV full-energy peak positions of the $\rm ^{137}Cs$ source using different packages, we found that the light collection efficiency of this type of package is $\rm 62\%$ that of a full coupling of the scintillator bottom surface. The optical grease is used for the optical coupling between the scintillator and SiPM array in the first payload. It is replaced with a silicon sheet in the second payload because of the fixed shape and good light collection efficiency of the silicon sheet.

\subsection{SiPM}
The SiPM has numerous attractive features, such as small size, insensitivity to magnetic fields, low power consumption, and light weight, which are crucial in space mission applications\cite{bib:8}, particularly nanosatellites. Therefore, four SiPM arrays are utilized as photoelectric converters. The SiPMs are the J-60035 type manufactured by SenSL and have a high photon detection efficiency (PDE) curve that matches reasonably with the emission spectrum of GAGG:Ce with a low dark count rate. Each SiPM chip has dimensions of $\rm 6.13\ mm\times 6.13\ mm\times 0.6\ mm$ and consists of 22,292 single-photon avalanche diodes with a microcell fill factor of 75$\rm \%$. The bias voltage supplied to the SiPM is 28.5 V, which is considerably lower than that required by a PMT. To reduce the dark noise and output capacitance of the SiPM array and manage the costs, a $\rm 4\times4$ SiPM array is adopted with a $\rm 2.45\ cm\times2.45\ cm$ area, which covers approximately one-third of the GAGG:Ce scintillator bottom surface area , as mentioned previously.

\begin{figure}[!htb]
	\includegraphics[width=0.45\hsize]{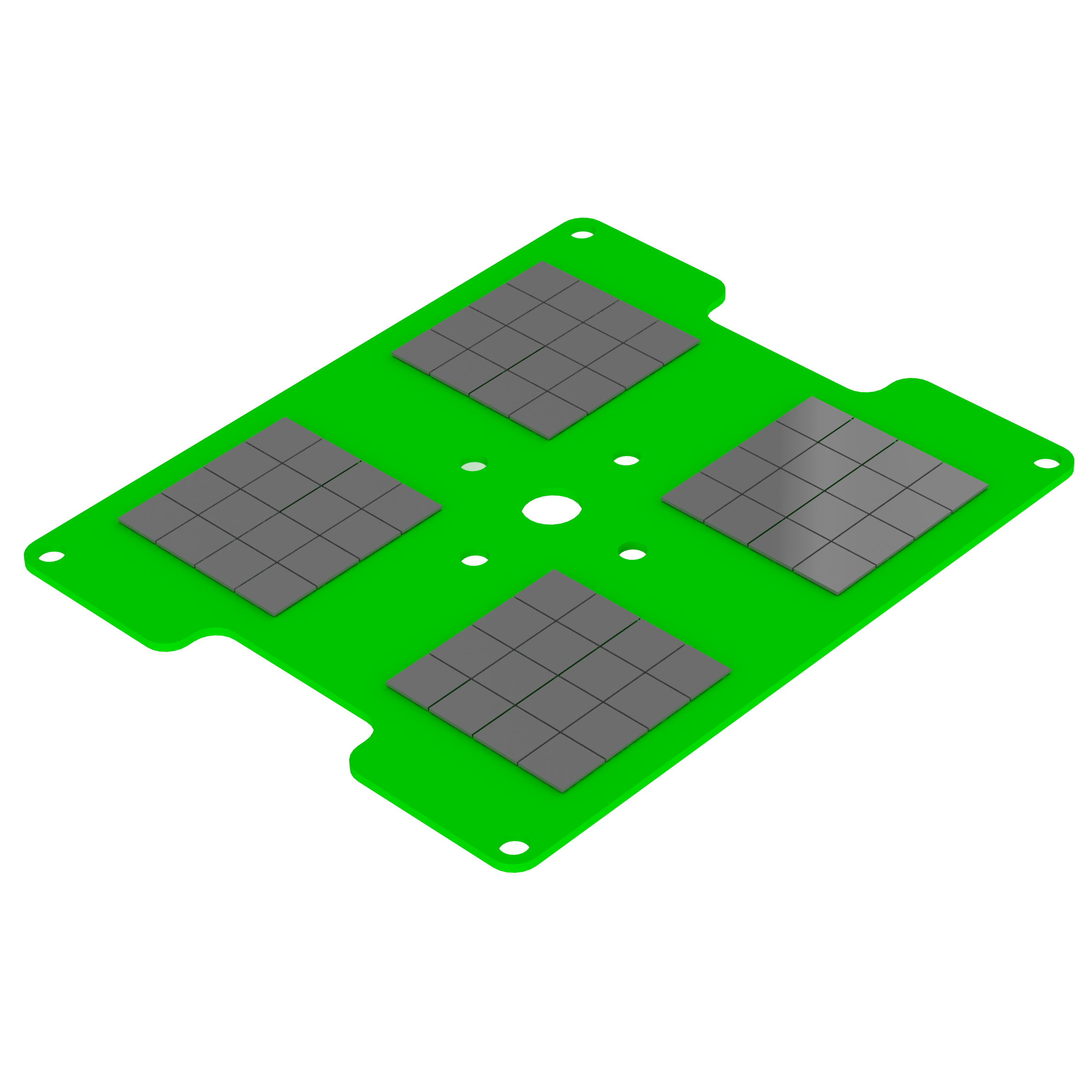}
	\includegraphics[width=0.45\hsize]{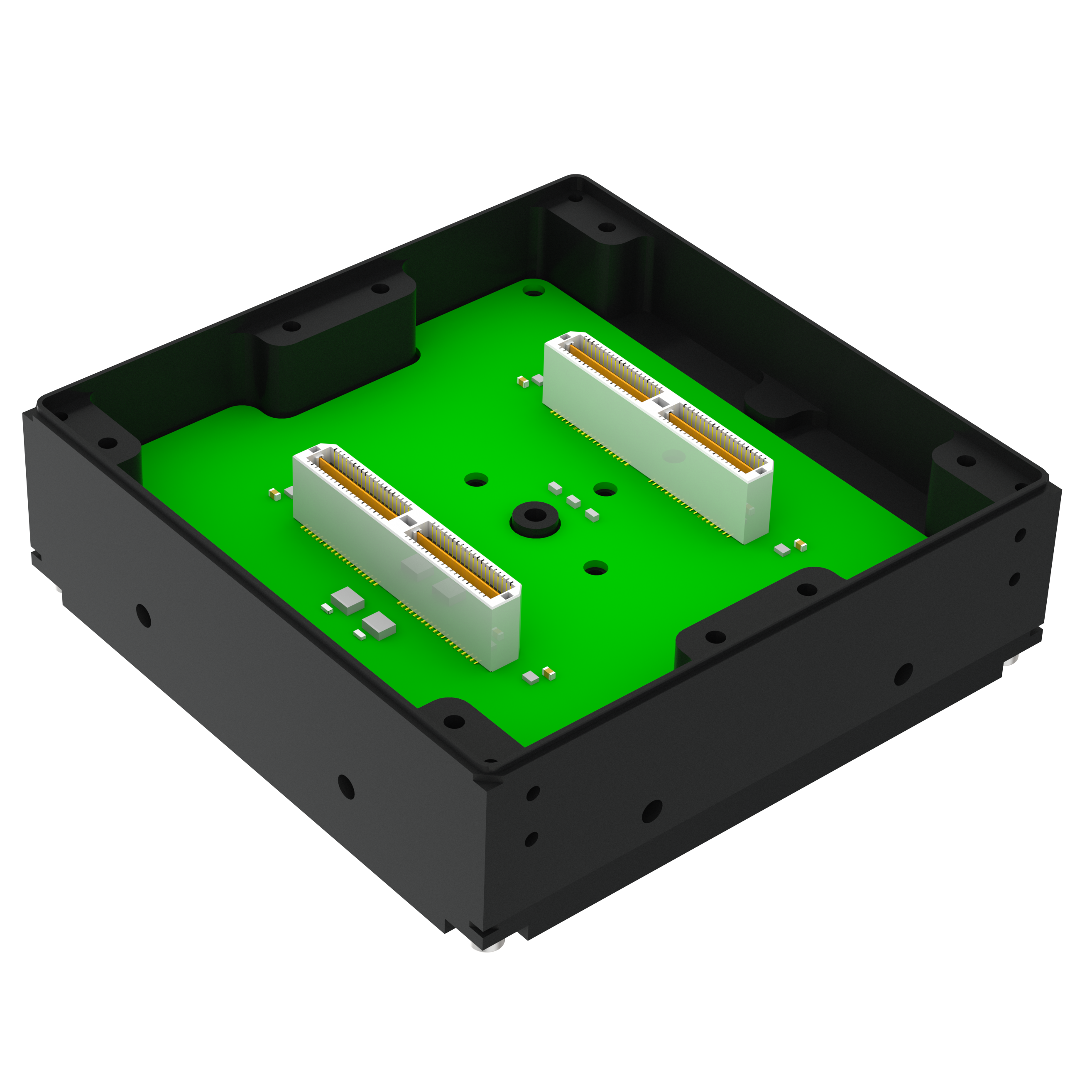}
	\caption{Top view (left) and bottom view (right) of the SiPM array board}
\label{fig:SiPM}
\end{figure}

Four SiPM arrays are integrated on a single board designed and manufactured in-house (Fig.\ref{fig:SiPM}). In addition, an SiPM through silicon via (TSV) package ensures that the SiPM fill factor of the printed circuit board (PCB) footprint is over $\rm 93\%$. Each array is powered independently, and the fast and standard outputs of every single SiPM chip are independently extracted to the FEE through a high-density connector $\rm QTE\_040\_03\_F\_D\_A$ manufactured by SamTec. Because the breakdown voltage of the SiPM changes with temperature, which affects the gain of the SiPM, there is a temperature-monitoring chip on the other side of the PCB for the correction of the SiPM gain.

\subsection{FEE}
\begin{figure}[!htb]
\includegraphics[width=0.9\hsize]{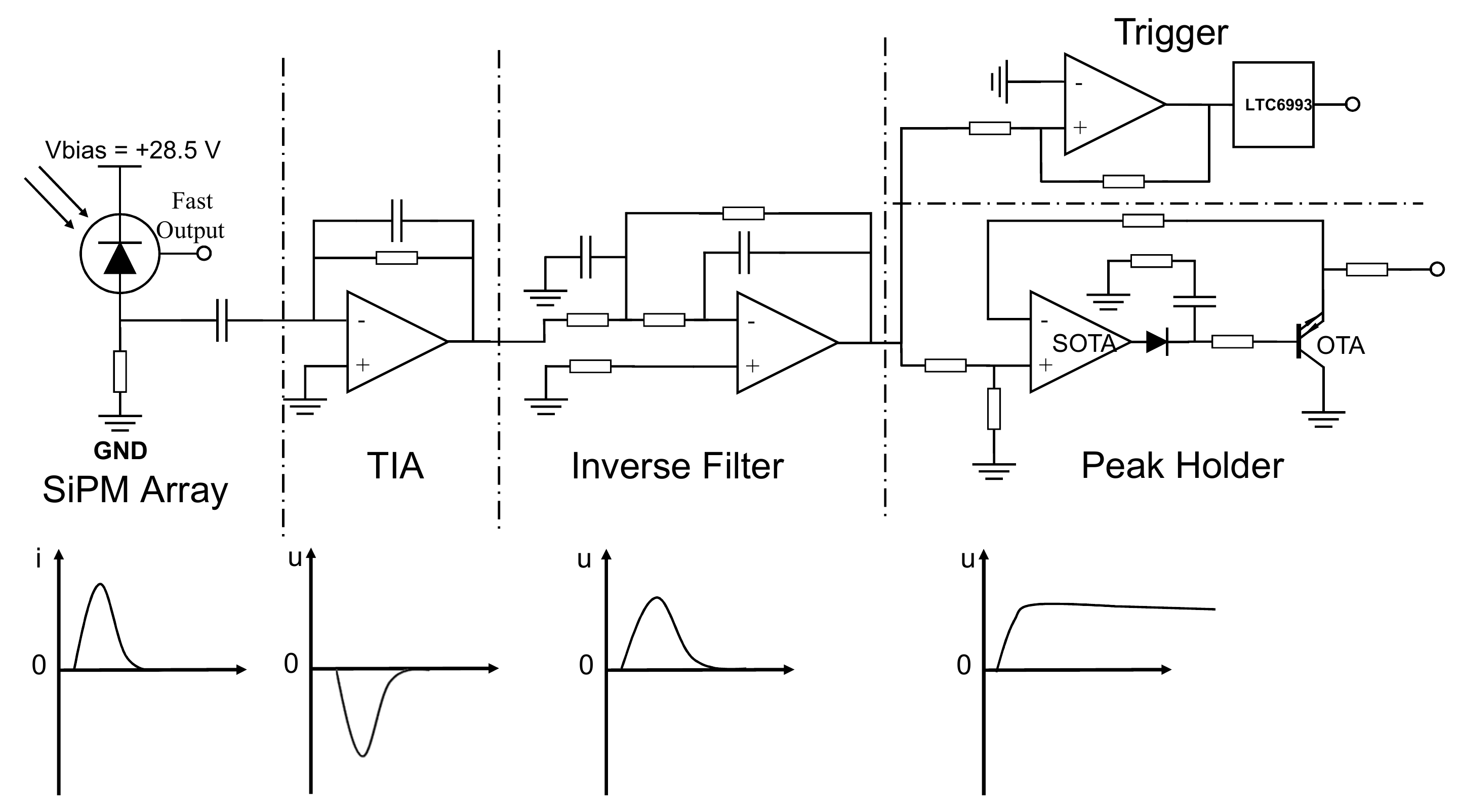}
\caption{Functional block diagram and output pulse shape of the FEE for one GRID channel}
\label{fig:FEE_SCH}
\end{figure}

A schematic of the FEE is shown in Fig.\ref{fig:FEE_SCH}. The standard output signals from one $\rm 4\times4$ SiPM array are directly shorted on the FEE and fed to a transimpedance amplifier (TIA) via an alternating current (AC) coupling, while the fast output signal pins are left floating. A $\rm 2\ k\Omega$ resistor connects the standard output of the array to the ground as the direct-current (DC) path of the standard output, which can restrain the current of the SiPM array to improve the system robustness. A standard high-speed amplifier OPA656 was adopted as the TIA amplifier, and the parameters of the TIA were optimized by detector modeling, as described in the next section. The TIA is followed by a low-pass filter circuit to reverse the signal and adjust its amplitude. The filter output is directed into two paths, which are connected to the trigger and peak hold circuits. The trigger circuit comprises a hysteresis comparator with an adjustable threshold and monostable pulse generator LTC6993, generating a high-level trigger signal for 2 $\mu$s without a retrigger. The high-bandwidth peak hold circuit comprises an operational transconductance amplifier (OTA) OPA615 and electronically controlled analog switch for discharge. Then, the four triggers and four peak hold signals are fed to the DAQ through a standard 2.54 mm connector.

\subsection{Detector optimization}
For such a compact detector, considerable efforts have been invested to improve the SNR, such as the scintillator reflector design, SiPM array design, and high-speed amplifier application. However, we do not have flexibility in terms of the scintillator and SiPM, whereas the TIA parameters significantly affect the SNR and can be analyzed and optimized in detail.

The transient response of the GAGG:Ce scintillator can be expressed as a single exponential decay signal, and its normalized transfer function is
\begin{equation}
\label{eq:1}
H_{{\rm G}}(s) = 	\frac{1}{s \times \tau_{{\rm GAGG}}+1}
\end{equation}
where $\tau_{\rm{GAGG}} = 100\ \rm{ns}$ is the GAGG:Ce decay time.

An accurate electrical model of SiPM is complex. However, neglecting the equivalent input resistance of FEE and quench capacitance can simplify the SiPM transient response to a single exponential decay signal, with a normalized transfer function as follows:
\begin{equation}
\label{eq:2}
H_{{\rm S}}(s) = 	\frac{1}{s \times \tau_{{\rm SiPM}}+1}
\end{equation}
where $\tau_{\rm{SiPM}} \approx\ 38\ \rm{ns}$ is the recovery time of the SenSL J-series SiPM\cite{bib:10}\cite{bib:11}.

With proper selection of feedback capacitance and resistance, the TIA can be treated as a second-order Butterworth filter with the following transfer function 
\begin{equation}
\label{eq:3}
H_{{\rm T}}(s, R_{\rm F}) = \frac{\Omega^2R_{\rm F}}{s^2+\frac{\Omega}{Q}s+\Omega^2}
\end{equation}
where $\Omega=2{\rm \pi} F$, $F=\sqrt{GBP/(2{\rm \pi} R_{\rm F}C_{\rm D})}$ is the approximate -3 dB bandwidth of the TIA circuit, $GBP$ is the gain bandwidth product of OPA656, $R_{\rm F}$ is the feedback resistance, and $C_{\rm D}$ is the output capacitance of the SiPM array. Here, $Q = 0.707$ is the quality factor of the Butterworth filter. Thus, the output pulse waveform of the TIA can be expressed as
\begin{widetext}
\begin{equation}
\label{eq:4}
Pulse(t) = E\times LY \times CE \times PDE \times e \times G \times \mathscr{L}^{-1}[H_{{\rm G}}(s)H_{{\rm S}}(s)H_{{\rm T}}(s)](t)
\end{equation}
\end{widetext}
where $E$ is the incident photon energy, $LY$ is the light yield of GAGG:Ce, $CE$ is the scintillation light collection efficiency determined by the scintillator and its packaging, $PDE$ is the PDE of the SiPM array, $e$ is the elementary charge, and $G$ is the gain of the SiPM.

The equivalent output noise voltage of the TIA contributed by the SiPM dark counts can be treated as a random pulse train. The standard deviation of the dark count noise can be obtained from Campbell's theorem\cite{bib:12}
\begin{equation}
\label{eq:5}
Vn_{\rm SiPM} = \sqrt{\bar n\int_{-\infty}^{+\infty}h^2(t)}
\end{equation}
where $\bar n$ is the dark count rate, $\bar n \approx {\rm 90\ kHz/mm^2} \times {\rm 576\ mm^2}$ for one SiPM array at $\rm 20\ ^\circ C$ with an operation voltage of 28.5 V, and $h(t)$ is the dark count pulse. The dark count pulse can be considered a Dirac delta pulse $e G \delta(t)$ convoluted by the SiPM and TIA response function; thus, 
\begin{equation}
\label{eq:6}
h(t) = e G \mathscr{L}^{-1}[H_{\rm S}(s)H_{\rm T}(s)](t).
\end{equation}
Considering that the output noise of the TIA is bandlimited, the equivalent output noise voltage contributed by the TIA can be estimated by a simple expression
\begin{widetext}
\begin{equation}
\label{eq:7}
Vn_{\rm TIA} = \sqrt{((I_{\rm N}R_{\rm F})^2+4kTR_{\rm F}+E_{\rm N}^2+\frac{(E_{\rm N}2\pi C_{\rm D}R_{\rm F}F_0)^2}{3})\times F_0}
\end{equation}
\end{widetext}
where $I_{\rm N}$ and $E_{\rm N}$ are the input current and voltage noise of OPA656, respectively, $4kTR_{\rm F}$ is the thermal noise of the feedback resistor, and $F_0 \approx 1\ {\rm MHz} $ is the band-limiting frequency of the TIA and low-pass filter system.
Therefore, the SNR can be defined as 
\begin{equation}
\label{eq:8}
SNR = \frac{{\rm max}(Pulse(t))}{\sqrt{Vn_{\rm SiPM}^2+Vn_{\rm TIA}^2}}.
\end{equation}

\begin{figure}[!htb]
\includegraphics[width=0.9\hsize]{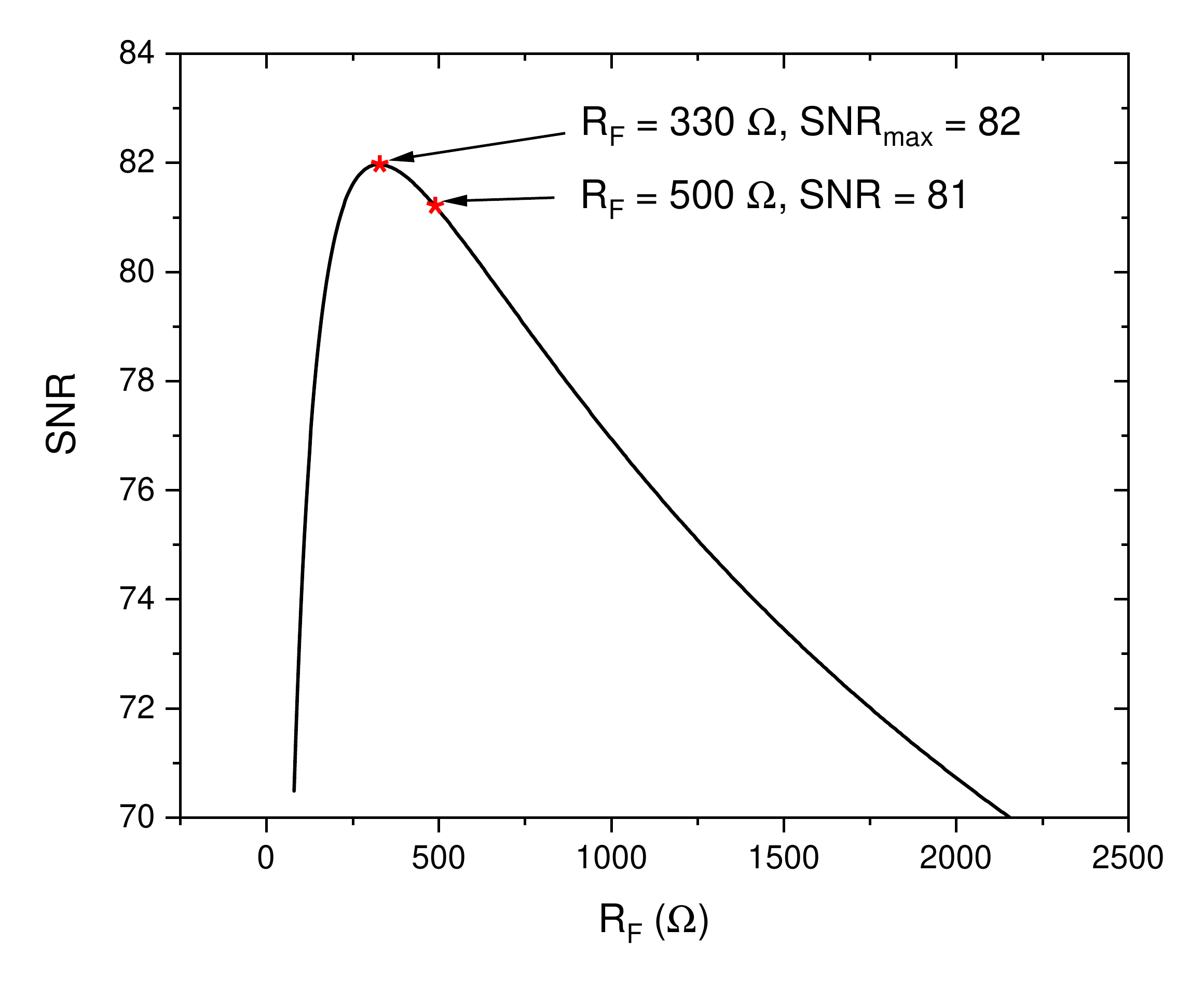}
\caption{SNR variation according to $R_{\rm F}$ with an incident photon energy of 59.5 keV at $\rm 20\ ^\circ C$, SiPM operating voltage of 28.5 V, and scintillator light yield of 32 ph/keV. The maximum SNR value is 82 at $R_{\rm F} = 330\ \Omega$.}
\label{fig:SNR_RF}
\end{figure}

The SNR varies with the feedback resistance $R_{\rm F}$, as shown in Fig.\ref{fig:SNR_RF}. $R_{\rm F}$ was set to $500\ \Omega$ to satisfy both the SNR and appropriate gain.

\section{Data acquisition and processing}
\subsection{Data acquisition electronics}

\begin{figure*}[!htb]
\includegraphics[width=0.9\hsize]{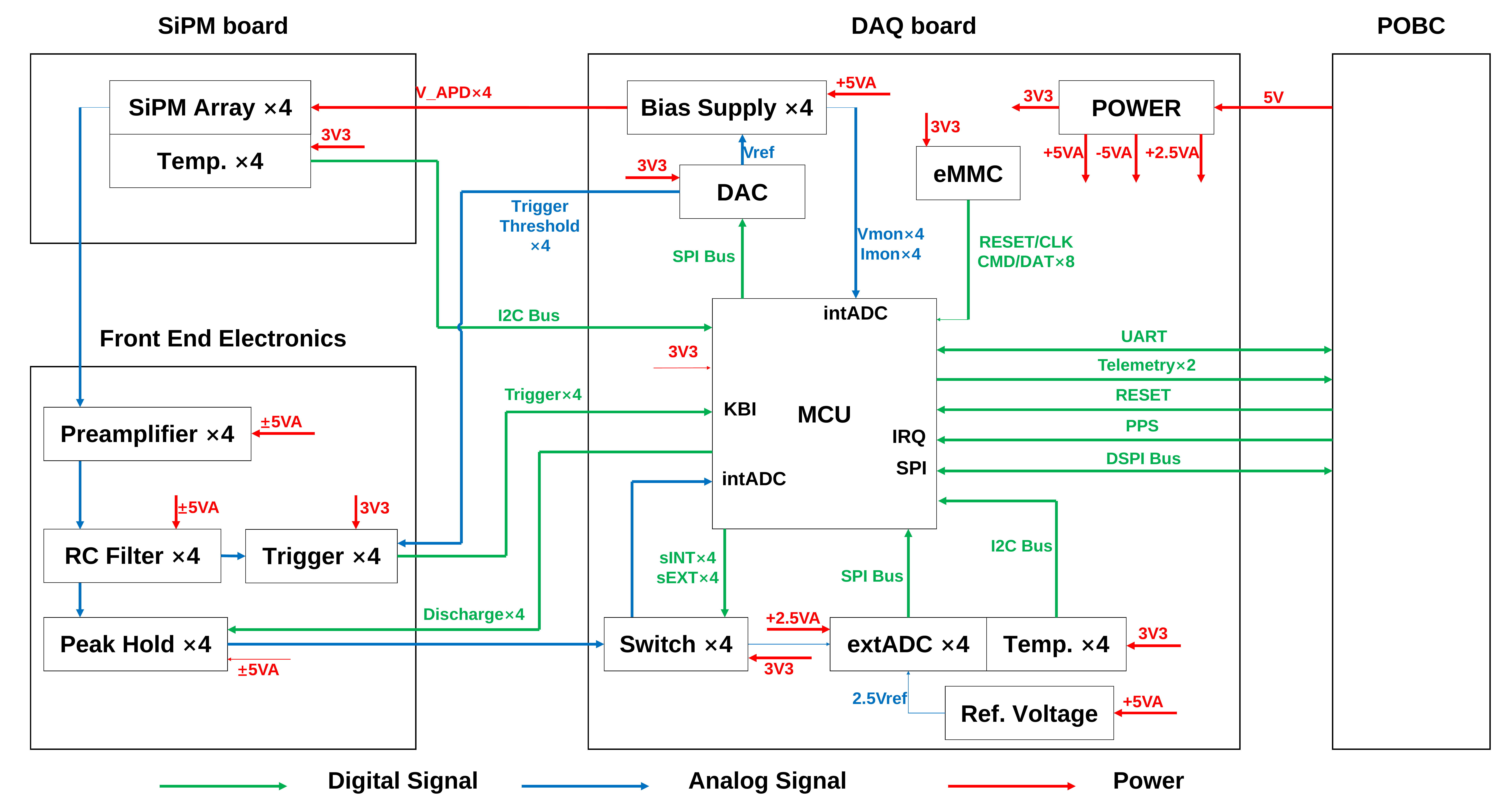}
\caption{Functional block diagram of DAQ and its connection with FEE, SiPM Carrier, and payload on-board computer board (POBC). The red lines represent the power bus, green lines represent digital signals, and blue lines represent analog signals. The debug interface and TTL-LVDS converter are not shown.}
\label{fig:DAQ_Block}
\end{figure*}
The DAQ board comprises the power regulator, analogue-digital converters (ADCs), MCU, embedded multimedia card (eMMC), and communication interface to process signals from the FEE, supply electricity for all analog and digital devices, supply and control bias voltage of SiPM arrays, process commands, format data for storage, and transmit data to the spacecraft, as shown in Fig.\ref{fig:DAQ_Block}. The power system regulates the +5 V input voltage to ±5 V for the analog devices, +2.5 V for the ADCs, and +3.3 V for the digital devices. An adjustable bias voltage of 0--40 V for the SiPM arrays can be generated through the SiPM bias voltage supply module, which can also monitor the bias voltage and leakage current. The DAQ core is a 32-bit ARM Cortex M0+ MCU KEA128 with 16 kB static random-access memory (SRAM) and 128 kB flash manufactured by NXP, which is an automotive-level MCU optimized for cost-sensitive applications and focuses on exceptional electromagnetic compatibility (EMC) and electrostatic discharge (ESD) robustness. A 512 MB single-layer cell (SLC) eMMC stores the raw science and housekeeping data with high reliability. The eMMC can store approximately 12 h of data based on existing data format definitions and background count rates of 500 counts/s. Peak hold signals from the FEE are sampled by four individual ADCs with 1M sample rate and 16-bit precision. The four internal ADCs of the MCU are alternatives of each other and can be selected through a gating switch for redundancy.

As shown in Fig.\ref{fig:DAQ_Block}, the DAQ electronics provides an interface with spacecrafts, which comprises a data bus using differential SPI protocol with LVDS level for the raw science and housekeeping data transmission, a UART interface for firmware update, a PPS interface for time calibration, and some general-purpose input/output (GPIO) interface for MCU reset, data request, boot configuration, and burst trigger (denoted as Telemetry and RESET).

\subsection{Flight firmware}
The flight firmware operates on the MCU without an operating system and comprises two parts: the boot loader and application program, residing in the internal flash memory of the MCU. The firmware provides limited online data processing capacity because of the low MCU performance. However, the configuration of all hardware, response to the triggers, command reception and processing, data storage and transmission, control and monitoring of the detector, and the application program update are provided, which can satisfy all necessary on-orbit requirements.

After the MCU is powered on, it  first runs the boot loader to check the Config$\rm \&$Trigger pin level, and updates the application program if high, or jumps to the application program if low. 

The application program is interrupt driven. Interrupts are generated on the following events:
\begin{itemize}
\setlength{\itemsep}{0pt}
\setlength{\parsep}{0pt}
\setlength{\parskip}{0pt}
\item [(1)] FEE trigger;
\item [(2)] timer interruption per second to record housekeeping data;
\item [(3)] PPS from the GPS module;
\item [(4)] spacecraft commands.
\end{itemize}

When the FEE trigger signals interrupt the MCU, the peak hold signals are sampled by four individual ADCs or four internal ADCs of the MCU, and the internal clock, PPS count, and UCT time will be recorded. When the converter sampling is finished, the MCU discharges the peak holder. Four channels work in a single thread in the present configuration, so other channels are in dead time, while one channel is triggered. The peak values and time information are stored in the eMMC. Every 1 s, the MCU records the housekeeping data, as listed in Table~\ref{table:TelemetryData}. The internal clock is recorded, while the PPS triggers the MCU for accurate time reconstruction of incident photons.

Currently, the application program processes the following commands from the spacecraft.
\begin{itemize}
\setlength{\itemsep}{0pt}
\setlength{\parsep}{0pt}
\setlength{\parskip}{0pt}
\item [(1)] Test communication status.
\item [(2)] Set the bias voltage of each SiPM array and trigger threshold for each channel.
\item [(3)] Process all modules self-test.
\item [(4)] Update the Coordinated Universal Time (UTC).
\item [(5)] Read data from eMMC.
\item [(6)] Read the housekeeping data.
\item [(7)] Set the ADC chosen switch.
\item [(8)] Erase the eMMC.
\end{itemize}

In daily operation, the instruction sequence will be sent to the spacecraft from the Earth station. Then, the spacecraft adjusts its attitude , powers on the detector, and controls the detector to set the SiPM bias voltage starting the observation, and powers off the detector after a specific observation time. The data stored in the detector eMMC will be read to the POBC's eMMC and downloaded at the appropriate time. 

\subsection{Data format}
The DAQ produces two types of data packets: the raw science and housekeeping data. All the raw data are stored  in the eMMC as a series of 512-byte packages. The definitions of housekeeping and raw science data are summarized in Tables~\ref{table:TelemetryData} and~\ref{table:ScienceData}, respectively. In a 512-byte raw science data package, the first trigger event occupies 3--25 bytes, and the other 43 trigger events occupy 26--499 bytes in the same format as shown, from 26 to 36 bytes.

\begin{table}[!htb]
\centering
 \caption{Housekeeping data package definition.}
 \label{table:TelemetryData}
 \begin{tabular*}{8cm} {@{\extracolsep{\fill} } llll}
 \toprule
 Bytes & Content & Bytes & Content \\
 \midrule
 0--2 & Head & 55--56 & TEMP of MCU \\
 3--6 & UTC & 57--64 & PPS to UTC \\
 7--14 & PPS Count & 65--72 & Internal clock to PPS \\
 15--22 & Internal clock & 73--492 & Telemetry dataset $\times$ 6 \\
 23--30 & TEMP of SiPM $\times$ 4 & 493--495 & Tail \\
 31--38 & TEMP of DAQ & 496--497 & CRC \\
 39--46 & Bias voltage $\times$ 4 & 498--511 & Vacancy\\
 47--54 & Leakage current $\times$ 4 & & \\
 \bottomrule
 \end{tabular*}
\end{table}

\begin{table}[!htb]
\centering
 \caption{Raw science data package definition.}
 \label{table:ScienceData}
 \begin{tabular*}{8cm} {@{\extracolsep{\fill} } llll}
 \toprule
 Bytes & Content & Bytes & Content \\
 \midrule
 0--2 & Head & 26 & Channel \\
 3 & Channel & 27--34 & Internal clock \\
 4--11 & Internal clock & 35--36 & ADC value \\
 12--13 & ADC value & 37--499 & Incident dataset $\times$ 42 \\
 14--17 & UTC & 499--501 & Tail \\
 18--25 & PPS count & 502--503 & CRC \\
 & & 504--511 & Vacancy \\
 \bottomrule
 \end{tabular*}
\end{table}

In a 512-byte housekeeping data package, there are seven housekeeping datasets occupying bytes 3--492 in the same format as shown in bytes 3--72. ``UT'', ``PPS count'', and ``Internal clock'' are indicated when the housekeeping data package is recorded. ``PPS to UTC'' indicates the corresponding PPS count last time UTC is received and ``internal clock to PPS'' denotes the corresponding internal clock last time PPS is received. From these data, we can precisely correspond the MCU internal clock to the UTC time, which is significant for GRB triangulation.

\section{Performance}
The detector performance can be investigated through experimental calibration and simulation~\cite{bib:14}. The performance of the GRID, including the energy-channel relations at different temperatures and biases, space uniformity, energy resolution, effective areas, angular responses, and detector noise, was calibrated experimentally in detail on the ground, and the effective areas and angular responses were investigated using a Monte Carlo simulation. The detector was irradiated with collimated radioactive sources in the laboratory (from 14 keV to 1.4 MeV) for calibration. Because of the low number of radioactive source emission lines in the low energy range, calibration measurements from 10 to 160 keV were performed with X-ray radiometry at the National Institute of Metrology of China. The detailed calibration and simulation results will be described in a future study. Here, we summarize the key features of low-energy performance, energy resolution, and high-rate performance, as well as the temperature dependence of the GRID.

\subsection{Low-energy performance}
Based on the SNR expression derived previously, if the minimum SNR required is 6, which means that the peak value of the signal amplitude is six times the standard deviation of noise, the lower limit for the low-energy detection can be derived theoretically. The light yield of the GAGG:Ce is calibrated experimentally in-house. The GAGG:Ce and a $\rm LaBr_3:Ce$ scintillator with a known light yield are irradiated by a $\rm ^{241}Am$ source, respectively, and the same PMT and electronics are used for the scintillation light readout. Considering the quantum efficiencies of the PMT cathode, the light yield of GAGG:Ce is estimated to be 16 ph/keV by comparing the 59.5 keV peak positions. Under normal observation conditions, where the temperature is $\rm 20\ ^\circ C$ and SiPM operating voltage is 28.5 V with a scintillation light collection efficiency of 62$\rm \%$ and PDE of the SiPM array, defined by the PDE of a single SiPM multiplied by the array's fill factor, of approximately $\rm 27.9\%$, the lower limit of the detector is 13 keV. The light yield of the crystal is lower than that reported in a previous study \cite{bib:9}. The provider of the GAGG:Ce crystal has improved the crystal growth process. Therefore, a crystal with a higher light yield will be used in the next detectors, with the lower limit of the GRID expected to extend to 10 keV. In addition, with the development of low dark count rates \cite{bib:15} and high PDE SiPMs\cite{bib:16}, the low-energy performance of these types of detectors can be further improved.

\begin{figure}[htb!]
\includegraphics[width=1\hsize]{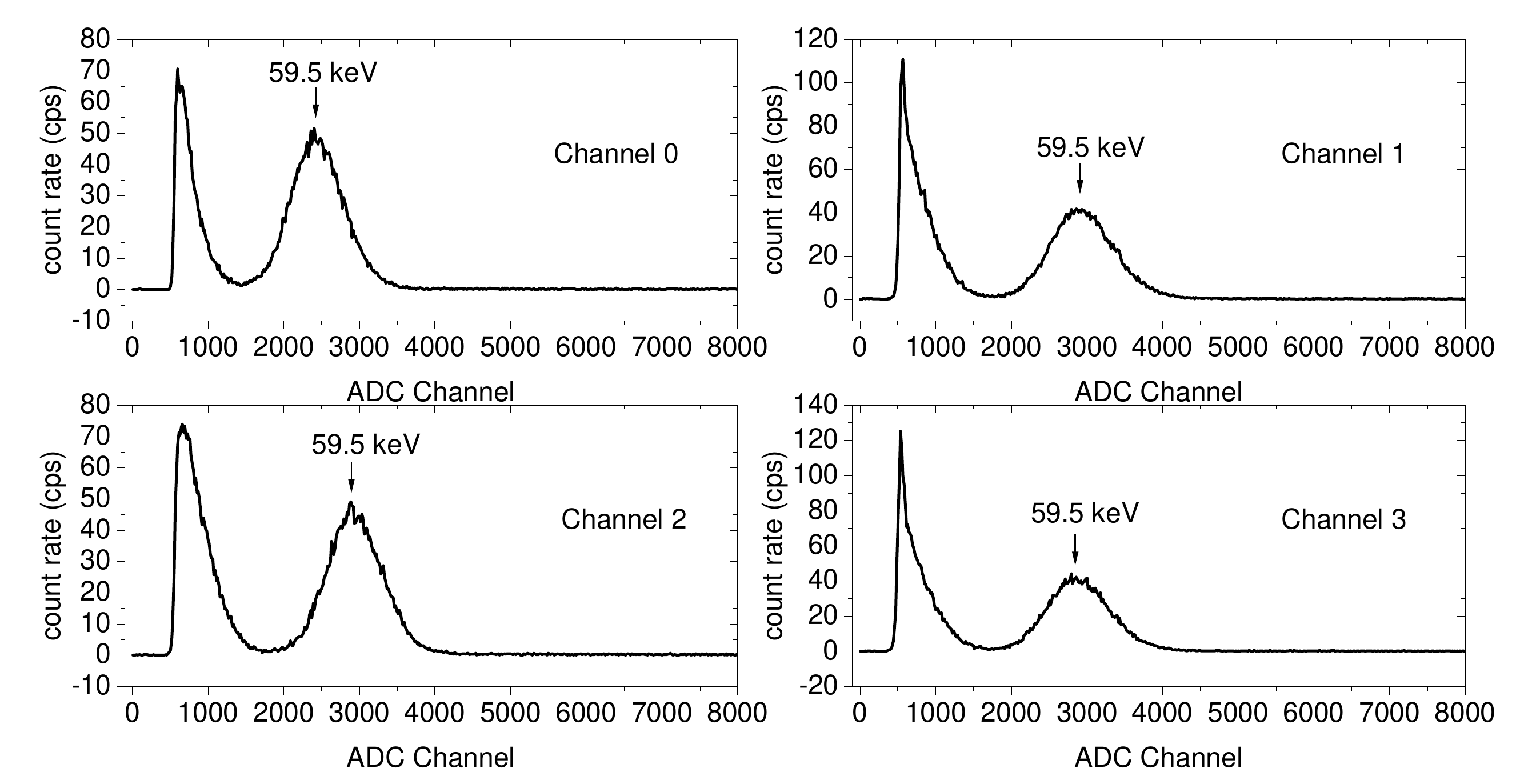}
\caption{Spectrum of $\rm ^{241}Am$ with a trigger threshold of 20 mV for four channels.}
\label{fig:Am241Spectrum}
\end{figure}

The trigger threshold was set to 20 mV (approximately 13 keV) and a spectrum of $\rm ^{241}Am$ with 59.5 and 13.5 keV X-rays was measured, as shown in Fig.\ref{fig:Am241Spectrum}. The performances of the four channels are not exactly the same owing to the difference in the scintillator light yield and light collection efficiency. However, the peak of the 59.5 keV X-ray can be observed clearly. The peak at 13.5 keV and dark count noise are mixed together near the threshold.

\subsection{Energy resolution}
\begin{figure}[!htb]
\includegraphics[width=0.9\hsize]{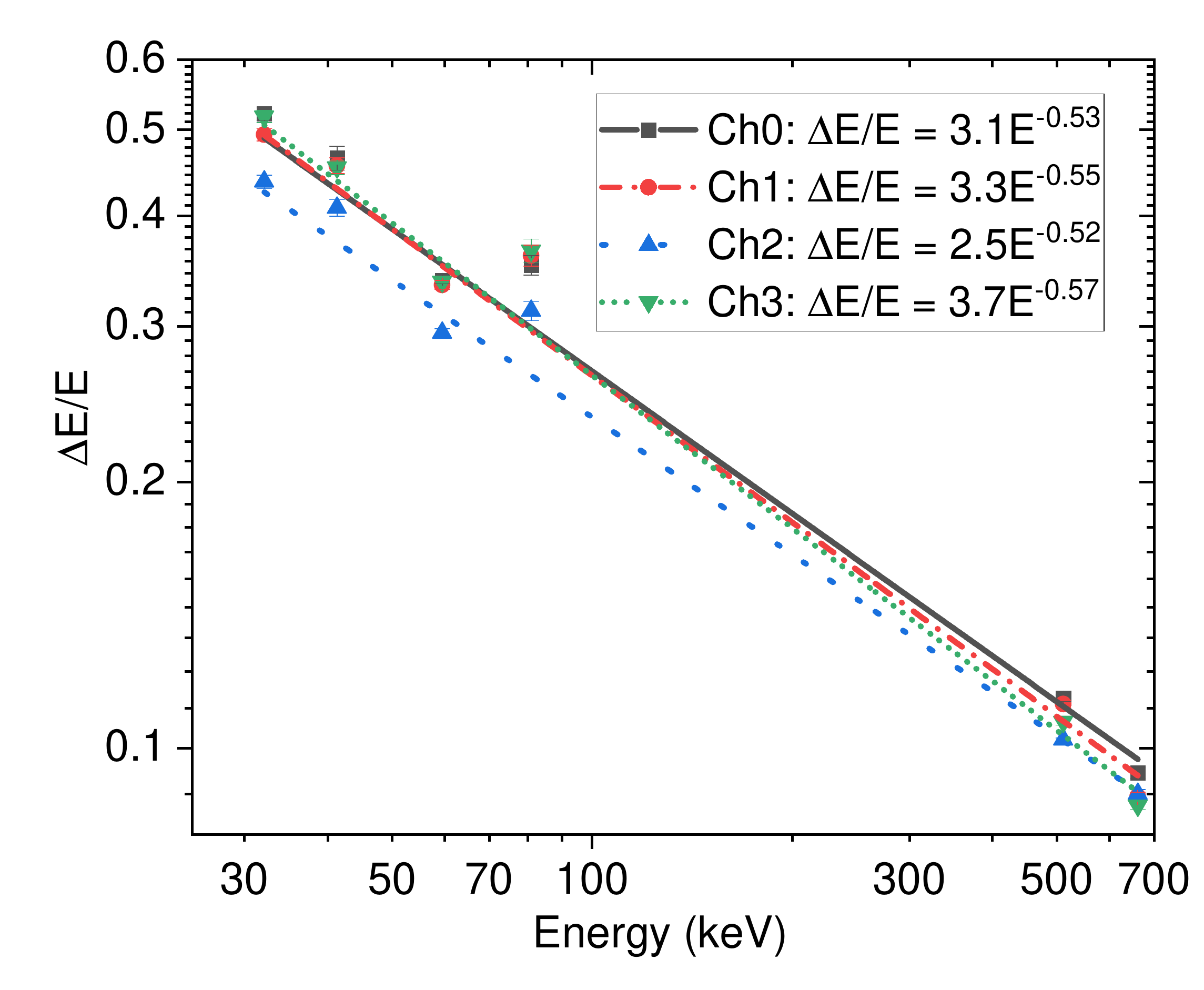}
\caption{Dependence of the detector energy resolution of four channels, and fitting curves and equations in terms of $y=ax^b$.}
\label{fig:EnergyResolution}
\end{figure}

The energy resolution results were calibrated by radioactive sources in the laboratory, which handled the detector and electronics noise, as well as the statistical fluctuation and energy nonlinearity. The sources used for calibration with their emitted photon energies are listed in Table~\ref{table:source}. As shown in Fig.\ref{fig:EnergyResolution}, the energy resolution $\frac{\Delta E}{E}$ is approximately proportional to $\frac{1}{\sqrt{E}}$ and is approximately $\rm 9\%$ at 662 keV. The poor fitting in the low-energy region is due to the nonlinearity of GAGG:Ce, particularly the distinct inconsistency at 81 keV, which is caused by the X-ray absorption edge of GAGG:Ce at approximately 70 keV\cite{bib:9}.

\begin{table}[!htb]
\caption{Radioactive sources used for calibration.}
\label{table:source}
\begin{tabular*}{5cm} {@{\extracolsep{\fill} } ll}
\toprule
Source & Energy (keV)\\ 
\midrule
$\rm ^{133}Ba$ & 32.1, 81.0 \\
\hline
$\rm ^{155}Eu$ & 41.3 \\
\hline
$\rm ^{241}Am$ & 59.5\\
\hline
$\rm ^{22}Na$ & 511\\
\hline
$\rm ^{137}Cs$ & 661.7 \\
\bottomrule
\end{tabular*}
\end{table}

\subsection{High-rate performance}
Two effects typically impair the performance of scintillation detectors at high photon rates: dead time and pulse pile-up. The pulse pile-up occurs when the count rate is so high that the pulses from successive events overlap in the FEE, which causes distortions in the measured spectrum that are difficult to characterize. These types of distortions are generally treated as systematic errors in the determination of the gamma-ray spectrum. Owing to the high bandwidth of the TIA and filter, the pulse of a single event lasts less than $\rm 1\ \mu s$ from generating a trigger to recover to baseline, which incurs little distortion in the measured GRB spectrum according to the relevant discussion about the Fermi GBM detector\cite{bib:13}. However, there is another type of pulse pile-up occurring at the peak-hold circuit; that is, a small signal will be overridden by a larger signal. This problem can be solved by a limit switch that restricts the peak-hold circuit input voltage to the ground level when the channel is triggered.

\begin{figure}[!htb]
\includegraphics[width=0.9\hsize]{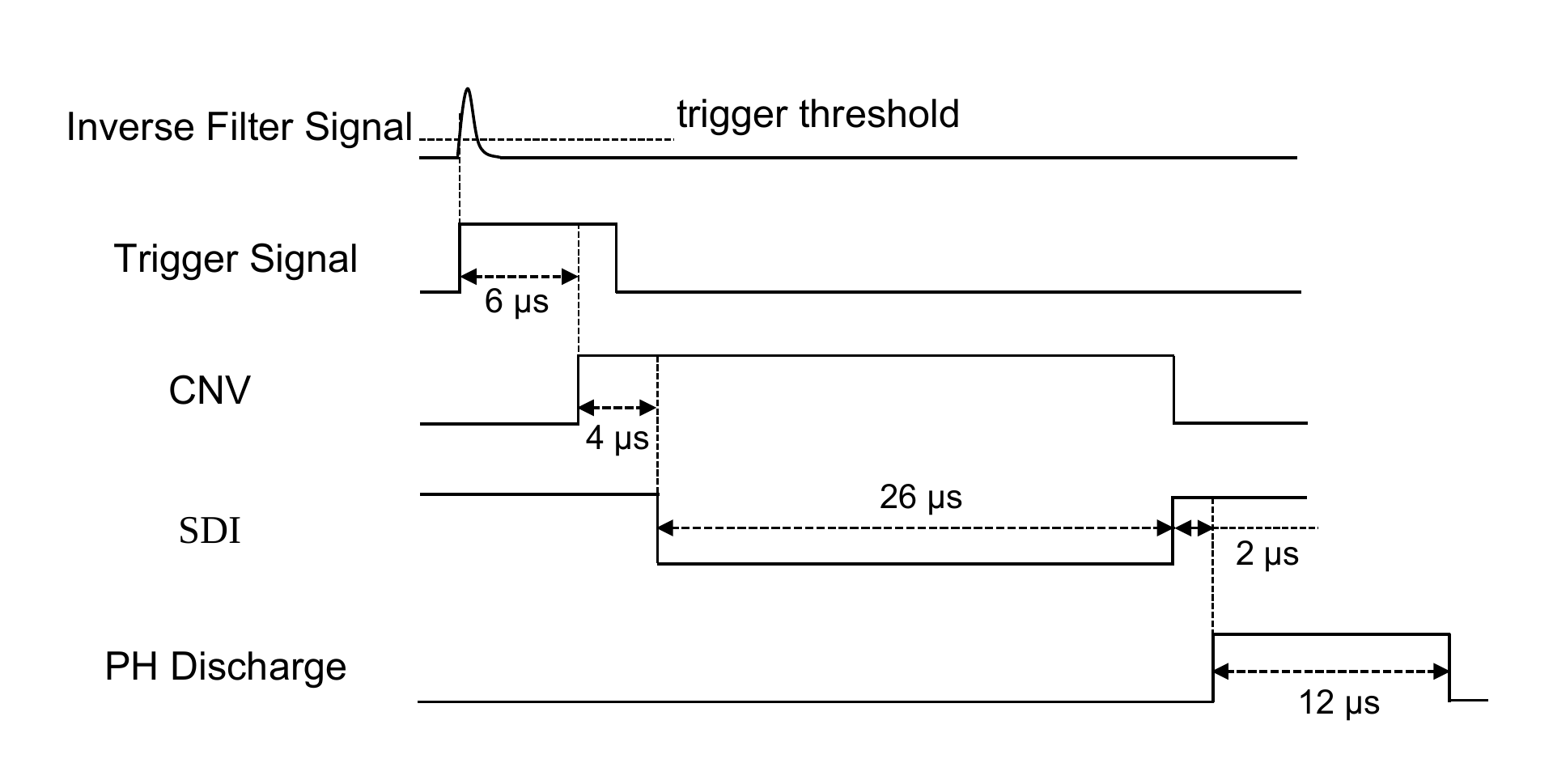}
\caption{Contribution of dead time in GRID. When the MCU is triggered by the FEE trigger output signal, a conversion input (CNV) is generated after 6 s to initiate the ADC conversions. The ADC conversion takes 4 $\mu$s, which is followed by a 26 $\mu$s MCU--ADC SPI communication time. Then, the discharge signal is generated after 2 $\mu$s and lasts for 12 $\mu$s until the ADC value is wrapped in the eMMC.}
\label{fig:deadtime}
\end{figure}

The nominal detector dead time is approximately 50 $\mu$s per event, which mainly consists of the MCU--ADC communication and MCU--eMMC communication, as shown in Fig.\ref{fig:deadtime}. However, in the last firmware version, which is used in the second GRID detector, the dead time is optimized to 15 $\mu$s with the same hardware design.

\subsection{Temperature dependence}
\begin{figure}[htb!]
\includegraphics[width=0.9\hsize]{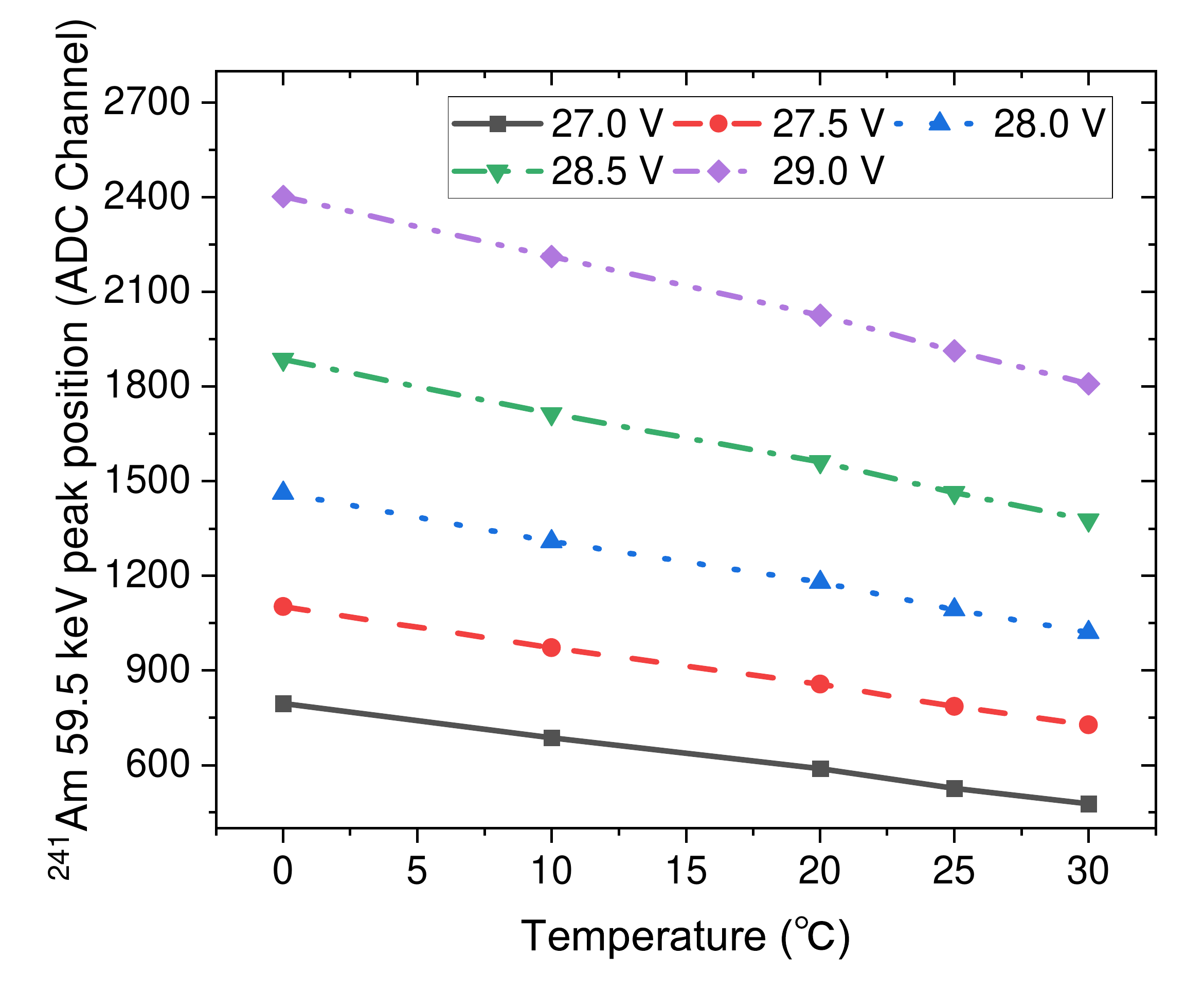}
\caption{Calibration results for the temperature dependence of the GRID detector channel 0 using $\rm ^{241}$Am source under bias voltages between 27.0 -- 29.0 V.}
\label{fig:TempBias}
\end{figure}

As the breakdown voltage of an SiPM varies with temperature, the gain of SiPM arrays is significantly affected by the temperature. The relationships between the gain, temperature, and bias voltages should be calibrated\cite{bib:17}. Figure~\ref{fig:TempBias} shows the calibration results for the temperature dependence of the GRID detector channel 0. The gain decreases with increasing temperature, and other channels show the same changing rule as channel 0. In a GRID detector, the bias voltage does not vary with temperature to stabilize the gain. The temperature of SiPM arrays will be recorded for offline correction of the energy-channel relation of the detector.

\section{Conclusion}
GRID mission is a student project with a dedicated and straightforward scientific goal: to detect and locate GRBs produced by neutron star mergers jointly with ground-based GW detectors in the local universe. In this paper, the detailed design of a GRID, electronics, and firmware, as well as the energy resolution and low-energy and high-rate performances of the detector are introduced. Further calibration of the detector, including the angular response, detection efficiency, and temperature response, using both simulations and experiments, will be reported in detail in a future study.

GRID mission was initially proposed and developed by students, with a considerable contribution from undergraduate students, and shall continue to operate as a student project in future. The current GRID collaboration involves more than 20 institutions and continues to grow. The purpose of GRID mission is twofold. In addition to its scientific goals, we hope to attract excellent students from different disciplines into astrophysics and train them to organize and participate in a multi-disciplinary collaboration, while learning how to build a real science project that covers hardware, data, and science.

In conclusion, GRID mission is a scientific collaboration that accepts students and scientists worldwide. Members can launch their own detectors, share the data, and produce scientific results under certain agreements. The detailed hardware and firmware design materials described in this paper will be part of a standard design package to be delivered and shared within the GRID collaboration community.


\begin{thebibliography}{99} 
\bibitem{bib:1} J. Wen, X. Long, X. Zheng, et al., GRID: a student project to monitor the transient gamma-ray sky in the multi-messenger astronomy era. Exp. Astron. 48, 77–95 (2019). \href{https://doi.org/10.1007/s10686-019-09636-w}{https://doi.org/10.1007/s10686-019-09636-w}
\bibitem{bib:8} V. Bindi, A. D. Guerra, G. Levi, L. Quadrani, C. Sbarra, Preliminary study of silicon photomultipliers for space missions. Nucl. Instrum. Methods Phys. Res. Sect. A. 572, 662–667 (2007). \href{https://doi.org/10.1016/j.nima.2006.12.011}{https://doi.org/10.1016/j.nima.2006.12.011}
\bibitem{bib:2} Chattopadhyay, T et al., BlackCAT CubeSat: a soft x-ray sky monitor, transient finder, and burst detector for high-energy and multimessenger astophysics. In: Proc. SPIE, Society of Photo-Optical Instrumentation Engineers (SPIE) Conference Series, vol. 10699, p. 106995S (2018). \href{https://doi.org/10.1117/12.2314274}{https://doi.org/10.1117/12.2314274}
\bibitem{bib:3} Racusin, J., Perkins, J.S et al., BurstCube: A CubeSat for Gravitational Wave Counterparts. arXiv:\href{1708.09292}{https://arxiv.org/abs/1708.09292} [astro-ph]. (2017).
\bibitem{bib:4} Ohno, M., Werner, N., Pal, A., et al., CAMELOT: Design and performance verification of the detector concept and localization capability. In: Proc. SPIE, Society of Photo-Optical Instrumentation
Engineers (SPIE) Conference Series, vol. 10699, p. 1069964 (2018). \href{https://doi.org/10.1117/12.2313228}{https://doi.org/10.1117/12.2313228}
\bibitem{bib:5} Pál, A., Ohno, M et al., GRBAlpha: a 1U CubeSat mission for validating timing-based gamma-ray burst localization. In: Proc. SPIE, Society of Photo-Optical Instrumentation
Engineers (SPIE) Conference Series, vol. 11444, p. 114444V (2020). \href{https://doi.org/10.1117/12.2561351}{https://doi.org/10.1117/12.2561351}
\bibitem{bib:6} Fuschino, F., Campana, R et al., HERMES: An ultra-wide band X and gamma-ray transient monitor on board a nano-satellite constellation. Nucl. Instrum. Methods Phys. Res. Sect. A. 936, 199–203 (2019). \href{https://doi.org/10.1016/j.nima.2018.11.072}{https://doi.org/10.1016/j.nima.2018.11.072}
\bibitem{bib:7} Braga, J., Durao, O.S.C et al., LECX: a cubesat experiment to detect and localize cosmic explosions in hard X-rays. Mon. Not. Roy. Astron. Soc. 493, 4852–4860 (2020). \href{https://doi.org/10.1093/mnras/staa500}{https://doi.org/10.1093/mnras/staa500}
\bibitem{bib:9} J. Iwanowska, L. Swiderski et al., Performance of cerium-doped Gd3Al2Ga3O12 (GAGG:Ce) scintillator in gamma-ray spectrometry. Nucl. Instrum. Methods Phys. Res. Sect. A. 712, 34–40 (2013). \href{https://doi.org/10.1016/j.nima.2013.01.064}{https://doi.org/10.1016/j.nima.2013.01.064}
\bibitem{bib:10} D. Marano, G. Bonanno et al., A New Accurate Analytical Expression for the SiPM Transient Response to Single Photons. In: Proc. 21st IEEE International Conference on Electronics, Circuits and System (IEEE, New York), pp. 514-517(2014). \href{https://doi.org/10.1109/ICECS.2014.7050035}{https://doi.org/10.1109/ICECS.2014.7050035}
\bibitem{bib:11} A. N. Otte, D. Garcia, T. Nguyen, D. Purushotham, Characterization of three high efficiency and blue sensitive silicon photomultipliers. Nucl. Instrum. Methods Phys. Res. Sect. A. 846, 106–125 (2017). \href{https://doi.org/10.1016/j.nima.2016.09.053}{https://doi.org/10.1016/j.nima.2016.09.053}
\bibitem{bib:12} Cohen, L., Generalization of Campbell’s theorem to nonstationary noise. In: Proc. 2014 22nd European Signal Processing Conference (EUSIPCO). p. 2415–2419 (2014)
\bibitem{bib:13} Meegan, Charles et al., The Fermi Gamma-Ray Burst Monitor. ApJ. 702, 791–804 (2009). \href{https://doi.org/10.1088/0004-637X/702/1/791}{https://doi.org/10.1088/0004-637X/702/1/791}
\bibitem{bib:14} Jiang, W., Yue C. et al., Comparison of Proton Shower Developments in the BGO Calorimeter of the Dark Matter Particle Explorer between GEANT4 and FLUKA Simulations, CHINESE PHYSICS LETTERS, 37, 119601 (2020). \href{https://doi.org/10.1088/0256-307X/37/11/119601}{https://doi.org/10.1088/0256-307X/37/11/119601}
\bibitem{bib:15} Haddadifam, T.,  Karami, M.A.. Dark count rate and band to band tunneling optimization for single photon avalanche diode topologies. Chin. Phys. B, 28, 458-464 (2019). \href{https://doi.org/10.1088/1674-1056/28/6/068502}{https://doi.org/10.1088/1674-1056/28/6/068502}
\bibitem{bib:16} Jia, J.Q., et al. EQR SiPM with P-on-N diode configuration. NUCL. SCI. TECH. 30, 19-25 (2019). \href{https://doi.org/10.1007/s41365-019-0644-9}{https://doi.org/10.1007/s41365-019-0644-9}
\bibitem{bib:17} Yu Sun, Zhi-Yu Sun et al., Temperature dependence of CsI:Tl coupled to a PIN photodiode and a silicon photomultiplier. NUCL. SCI. TECH. 30, 77-85 (2019). \href{https://doi.org/10.1007/s41365-019-0551-0}{https://doi.org/10.1007/s41365-019-0551-0}
\end{thebibliography}
\end{document}